\documentclass[a4paper,12pt]{article}
\usepackage[utf8]{inputenc}
\usepackage[affil-it]{authblk}
\usepackage{cite}
\usepackage{amssymb}
\usepackage{amsmath}
\usepackage{amsfonts} 
\usepackage{amsthm}
\usepackage{tikz}
\usepackage{tikz-qtree}
\usepackage{hyperref}

\theoremstyle{definition}
\newtheorem{definition}{Definition}[section]
\theoremstyle{plain}

\providecommand{\keywords}[1]{\textbf{Keywords: } #1}

\title{Publicly verifiable delegative democracy with secret voting power}
\author{Dimitrios Karoukis \thanks{Electronic address: \texttt{dim@deliberative.io}}}
\affil{Deliberative Technologies}

\begin{document}

\maketitle

\begin{abstract}

In a democratic setting, we introduce a commitment scheme which allows for transparent validation of transfers and reversible delegations of voting power between citizens without sacrificing their privacy.
A unit of voting power is publicly represented by the Merkle root of a tree consisting of its latest owner's public key, a random nonce and the Merkle root of the tree of its previous owner's public key and random nonce and so on.
A transition includes the input units, their owner's public keys and signatures, the hashes of their nonces and the output units generated with the new owners' public keys and random nonces.
In case of a delegation, the receiver provides the sender with the hashed random nonces and hashed public keys for the output units.
In case of a transfer, only the precomputed output units are provided by the receiver.
In a reversal, a historical owner reveals the hashes of the nonces and public keys that resulted in the subsequent units.
To vote, the owner reveals the actual nonces and public keys.

\end{abstract}


\section{Introduction}

Voter privacy, which tends to be a requirement for public and private institutions, is often one of the main impediments to the adoption of digital democracy tools.
While many of these tools can offer benefits such as more expressive voting paradigms than single-party ballots, increased citizen engagement, reduced public cost of voting and many more, they do require various levels of trust to be placed on operators who may not be entirely trustworthy to handle such sensitive information.
Democratic systems with digital components need to safeguard societies against the mishandling of those data.
Those safeguards should exist by design and they should be based on strong cryptographic guarantees and not rely on cheap promises made by corruptible individuals.

In our setting, the state of a democratic system is the voting power that every voter has.
We assume that there is a public ledger, the famous bulletin board, which records all the state transitions throughout the life cycle of the voting process, so as to serve as a common knowledge point of reference for all the participants who want to recreate the historical evolution of the state. In this setting, if a transition's initiator and receiver are both publicly visible, it is trivial to verify that one entity which transfers or delegates some of their voting power to another entity is actually in the possession of this voting power.

However, problems arise if we require publicly verifiability of transitions with privacy at the same time.
In our setting, privacy is achieved through \textit{forward secrecy}.
This means that, in all state transitions, there is no disclosure of who the receiver of the voting power is.
Public verifiability with privacy means that any external observer is able to verify that the initiator is indeed the latest owner of the voting power before executing the state transition and the receiver can verify that they are in possession of it. A byproduct of this property is that no third party can calculate all the available voting power that a citizen has at any given time. Another problem arises if we want the system to also support reversibility of delegations, because then the initiators need to be able to prove that they precede other owners in the chain of custody of the voting power without necessarily knowing the identity of all their predecessors and successors.

%

Our design was conceived due to the need for a
privacy-preserving mechanism that could accommodate a decentralized digital infrastructure for the voting paradigm by the name of \textit{liquid democracy} \cite{blum2016liquid} and its extensions \cite{zhang2021tracking}\cite{karoukis2021deliberative}.
In liquid democracy, voters can either
vote for themselves on a given issue or reversibly \textit{delegate} their voting power
to some other voters that they deem more capable of making an informed decision on a given subject.
To support this type of models without keeping track of the chain of custody of the voting power in the open was, to our knowledge, an open problem.

To solve the problem, we draw inspiration from a structure that is known as \textit{Merkle trees}\cite{merkle1987digital}, which are an example of a \textit{cryptographic commitment scheme}.
In a commitment scheme, there is a commitment algorithm and a reveal algorithm.
The commitment algorithm takes as input the message that a party wants to commit to and produces a commitment string, which is a fixed-size representation of the message.
The commitment string should be \textit{binding}, meaning that it should be computationally infeasible for the party to change the committed message without changing the commitment string,
and \textit{hiding}, meaning that the committed message should be computationally infeasible to determine from the commitment string.
The reveal algorithm takes as input the commitment string and the committed message, and produces a proof that the committed message corresponds to the commitment string.
The proof should convince anyone that the message was indeed committed to and that the commitment string was generated correctly.

In our scheme, initially, a \textit{unit of voting power} is represented by the \textit{Merkle root} of a tree consisting of a sufficiently large random nonce and a public key \cite{koblitz1987elliptic} whose secret counterpart the original owner controls.
After a transition, a unit is represented by the Merkle root of a tree consisting of a new random nonce, a subsequent owner's public key and the previous unit identifier.
A Merkle root is essentially the result of repeated hashing, where a hash is a collision-resistant, one-way function that provides a fixed-size, unique identifier of its input without revealing any information regarding its content.
If we assume a finite set of voters and common-knowledge public keys, it would be trivial to calculate all their public keys' hashes to find the owner of the voting power if there were no nonces, but it is computationally infeasible to also calculate the locally-generated, sufficiently large random nonces that need to be combined with the public keys to generate the specific Merkle roots.

The only individual who can know and verify that they are the owner of a unit is the individual who knows the plain text random nonce and the public key with its secret counterpart which, combined with the identifier of the unit before the transition that gave them possession of it, is sufficient to recreate the new identifier.
In a transfer, the receiver provides the sender with the hash of the concatenation of the two elements' hashes, which means that the sender cannot reveal the two hashes separately, therefore making the transfer irreversible. In a delegation, the two hashes are provided separately by the receiver, making the delegation reversible, in the sense that both of them can reveal the hashes but only the receiver can reveal the plain text elements.
To an outside spectator, all units are same-length random strings.
If a recorded transition only mentions the output units, no spectator can infer the identity of the receiver.

%
%
%
%
%

\section{Related work}

It would be natural to assume that privacy should be a desirable property of digital democracy systems, yet it has been discussed in \cite{nejadgholi2021short} and \cite{ford2020liquid} that, while ballot secrecy offers some \textit{coercion resistance}, it creates the pitfall that a vote receiver can mislead their delegators by secretly deviating from the action that was expected of them.
This is why it is suggested in \cite{ford2020liquid} that if a model uses the \textit{fake persona} pattern to induce anonymity and coercion resistance, which is a pattern that can also be used in our model through the issuing of multiple key pairs by the same voter, it should disallow the issuing of new identities by voting power receivers so that they can be held accountable for their actions by their delegators.
Our model's reversible delegations, combined with the fake persona pattern and forward secrecy, can accommodate for a vote tally design where the voters observe a preliminary result of the vote after all the delegations have happened and votes have been cast and then be given a chance to reverse their delegation once more if they feel that the result did not satisfy them, thereby implicitly holding their previous voting power receiver accountable for their potential mischief.

The most important design of a private, coercion resistant voting system in the literature is the JCJ/Civitas \cite{juels2005coercion} \cite{clarkson2008civitas} scheme, which is based on zero-knowledge proofs and random permutations of the votes. This introduced the notion of \textit{fake credentials} to ensure coercion resistance through the inability of the coercer to know which credential is fake and which is real. This principle has been improved upon in systems such as \cite{smyth2019athena} and it has been extended by \cite{spycher2012new} and \cite{kulyk2017coercion} to accommodate proxy voting by introducing \textit{delegation credentials with ranking}. With these items, a vote can be delegated multiple times, and, at the tally, only the delegation with the highest ranking will be counted.
This allows the voter to send a low ranking delegation to their coercer and use a higher ranking one later in order to reclaim their voting power.
Our design can accommodate this approach since multiple delegations can be executed within one recorded transition if the computed output unit identifier is the result of more than one evolutions of the same voting power unit.
This would assume of course an initial voting power distribution where no entity controls the original keys.
Another approach for coercion resistance has been the public display of evidence of coercion \cite{grewal2013caveat}.
For the sake of completeness, there also exists a class of models that address full secrecy but not coercion resistance, the so-called \textit{statement voting} scheme \cite{zhang2019statement}.

Some problems that may arise from zero-knowledge proof systems are the increasing size of the proofs as delegations evolve and the expensive computations that may need to be performed either by the user or by some centralized system. A solution to the issue of increasing proof size is the use of \textit{malleable proofs} \cite{chase2012malleable}, whose implementation in e-voting systems has been explored in \cite{bernhard2013towards}. These do not increase in size overtime and they retain their desirable properties.
In our commitment scheme, transfers and delegations have fixed-size proofs since they only mention elements of the latest evolution stage of the voting power unit in question and only the computed output units. In a reversal, however, the proof size may be larger if the historical owner has ranked credentials.
Moreover, the computation of a voting power unit identifier is efficient in time and system requirements and it can be performed by the users themselves, who are also responsible for exchanging the relevant data for a transfer or delegation between them in a \textit{peer-to-peer} fashion.

Another problem that arises in the fake credentials paradigm due to ballot secrecy in liquid democracy is the problem of \textit{delegation cycles}, which is the situation where one voter delegates their vote to another and after some number of redelegations it returns to the original voter unexpectedly.
A centralized solution to this problem would be, according to \cite{nejadgholi2021short}, an \textit{oracle} to which all the votes are sent and that can detect such cycles and remove the inducing transitions. This would of course create problems for coercion resistance, it would reduce voter privacy without any serious upside and it would deteriorate the autonomy of a publicly verifiable decentralized system.
In our system there always exists some interaction between the sender and the receiver of the voting power, which eliminates the probability of an unexpected or an unaccountable delegation.

\section{Model}

Suppose that we have a society $N$ of $|N| < \infty$ individuals.
Each individual $i \in N$ is in possession of a mnemonic generated from a sufficiently large dictionary, which translates to a random master seed that they can use to generate multiple unique
\textit{cryptographic key pairs} $(pk_i^j, sk_i^j)$ with a \textit{public} part $pk_i^j$ that can be revealed and a \textit{secret} part $sk_i^j$ which is only known to $i$
\cite{maxwell2011deterministic}\cite{gutoski2015hierarchical}.

There is a \textit{set of states} $S$ with initial state $S_0 \in S$.
For each \textit{time slot} $t < \infty$,
the \textit{state of the world at} $t$, $S_t \in S$, is represented by a finite set of \textit{voting power units} $v_t^{i,j}$, which are the \textit{Merkle roots} of trees consisting of \textit{random nonces} $n_{t}$ and $pk_i^j$.

\begin{definition}
A hash function $H$ is a deterministic algorithm that takes an input message $\mu$ of arbitrary length and produces a fixed-size output $h = H(\mu)$.
The hash function satisfies the following properties:
\begin{itemize}
\item \textbf{Collision resistance}: It is computationally infeasible to find two distinct messages $\mu_1$ and $\mu_2$ such that $H(\mu_1) = H(\mu_2)$.
\item \textbf{Preimage resistance}: Given a hash value $h$, it is computationally infeasible to find a message $\mu$ such that $H(\mu) = h$.
\item \textbf{Second preimage resistance}: Given a message $\mu_1$, it is computationally infeasible to find another message $\mu_2$ such that $H(\mu_1) = H(\mu_2)$.
\end{itemize}
\end{definition}

\begin{definition}
A Merkle tree is a binary tree in which each leaf node contains a hash of a data block, and each non-leaf node contains a hash of the concatenation of its two child nodes. The Merkle root is the hash of the concatenation of the top two child nodes.
A visual representation of a Merkle tree and root with $L_1,\ldots,L_5$ leafs, $H(x)$ hash function and $||$ concatenation symbol can be seen in the following figure:
\end{definition}

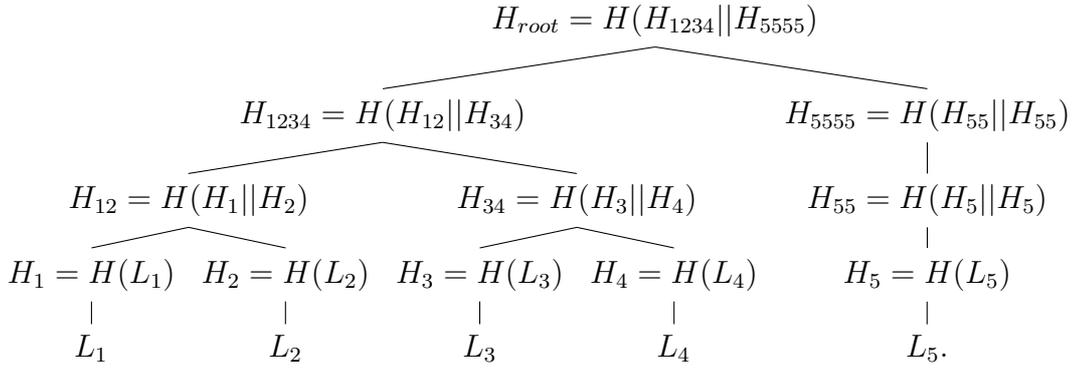
\begin{figure}[h!]
\centering
\begin{tikzpicture}
\tikzset{level 1/.style={level distance=36pt}}
\tikzset{level 2/.style={level distance=32pt}}
\tikzset{level 3+/.style={level distance=28pt}}
\Tree [
    .{$H_{root} = H(H_{1234} || H_{5555})$}
    [
        .{$H_{1234} = H(H_{12} || H_{34})$}
        [
             .{$H_{12} = H(H_1 || H_2)$}
            [
                .{$H_1 = H(L_1)$} {$L_1$}
            ]
            [
                .{$H_2 = H(L_2)$} {$L_2$}
            ]
        ]
        [
            .{$H_{34} = H(H_3 || H_4)$}
            [
                .{$H_3 = H(L_3)$} {$L_3$}
            ]
            [
                .{$H_4 = H(L_4)$} {$L_4$}
            ]
        ]
    ]
    [
        .{$H_{5555} = H(H_{55} || H_{55})$}
        [
            .{$H_{55} = H(H_5 || H_5)$}
            [
                .{$H_5 = H(L_5)$} {$L_5.$}
            ]
        ]
    ]
]
\end{tikzpicture}
\end{figure}

There are three types of state transitions that are supported in our design, the \textit{transfer}, the \textit{delegation} and the \textit{reversal}. The difference between the first two is that a delegation is \textit{reversible} while the transfer is not.
In this design, the receiver provides the sender with the arguments that will decide whether it is a reversible or irreversible transition.
With these arguments, the sender calculates locally the new voting power unit identifier, while keeping the knowledge of the actual components of the Merkle tree in the hands of the receiver. The receiver will be able to verify that the transition happened because they can also calculate locally the resulting voting power unit identifier.

\begin{definition}
  In this commitment scheme, the calculation of a value unit identifier from user $i \in N$ to $i' \in N$ with corresponding key indexes $K^i$, $K^{i'}$ where $j \in K^i$ and $j' \in K^{i'}$ and from state time $t$ to $t+1$ is represented by the following commitment algorithm:
  \begin{equation}\label{eq:unit}
    v_{t+1}^{i',j'} = H\left\{H\left[H\left(n_1\right) \Big|\Big| H\left(pk_{i'}^{j'}\right)\right] \Big|\Big| H\left(v_t^{i, j}\right)\right\}.
  \end{equation}

\end{definition}

To visualize the evolution of the identifier of a voting power unit through time, let's assume that a hypothetical unit's state of ownership mutates from one individual to another in each of 3 steps. This would create a graph like the following:

\begin{figure}[h!]
\begin{tikzpicture}[scale=0.73]
\tikzset{level 1/.style={level distance=45pt, sibling distance=-20pt}}
\tikzset{level 2/.style={level distance=45pt, sibling distance=5pt}}
\tikzset{level 3+/.style={level distance=35pt, sibling distance=5pt}}
\Tree
[
    .{$v_2^{i'',j''} = H\left(\tilde{v}_2^{i'',j''} || v_1^{i',j'}\right)$}
    [
        .{$\tilde{v}_2^{i'',j''} = H\left(h_{n_2} || h_{p_2}\right)$}
        [
            .{$h_{n_2} = H(n_2)$} {$n_2$}
        ]
        [
            .{$h_{p_2} = H(pk_{i''}^{j''})$} {$pk_{i''}^{j''}$}
        ]
    ]
    [
        .{$v_1^{i',j'} = H\left(\tilde{v}_1^{i',j'} || v_0^{i,j}\right)$}
        [
            .{$\tilde{v}_1^{i',j'} = H\left(h_{n_1} || h_{p_1}\right)$}
            [
                .{$h_{n_1} = H(n_1)$} {$n_1$}
            ]
            [
                .{$h_{p_1} = H(pk_{i'}^{j'})$} {$pk_{i'}^{j'}$}
            ]
        ]
        [
            .{$v_{0}^{i,j} = H\left(\tilde{v}_0^{i,j} || v_{-1}^{\tilde{i},\tilde{j}}\right)$}
            [
                .{$\tilde{v}_0^{i,j} = H\left(h_{n_0} || h_{p_0}\right)$}
                [
                    .{$h_{n_0} = H(n_0)$} {$n_0$}
                ]
                [
                    .{$h_{p_0} = H(pk_i^j)$} {$pk_i^j$}
                ]
            ]
            [
                .{$v_{-1}^{\tilde{i},\tilde{j}} = H\left(h_0 || h_0\right)$}
                [
                    .{$h_0 = H(0)$} {$0$}
                ]
            ]
        ]
    ]
]
\end{tikzpicture}
\caption{The evolution of a voting power unit from $v_0^{i,j}$ to $v_1^{i', j'}$ to $v_2^{i^{''}, j^{''}}$.}
\label{fig:unit}
\end{figure}
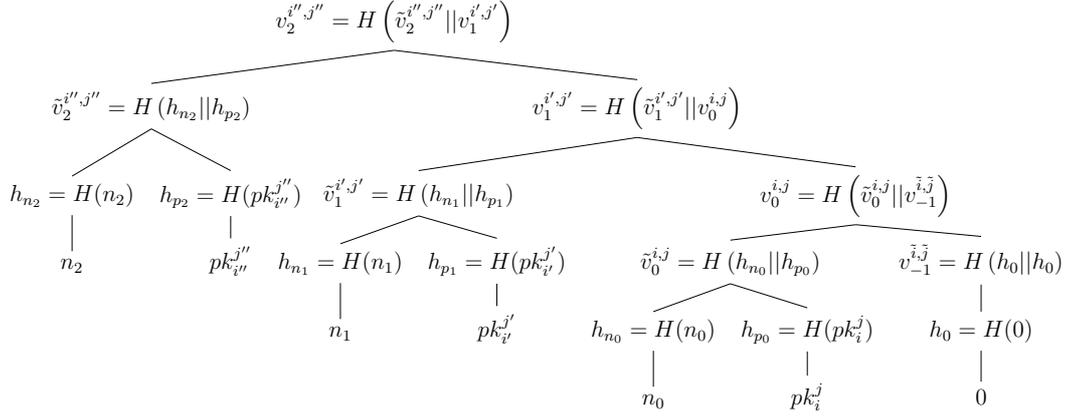

\begin{definition}
A transfer or a delegation of a voting power unit $v_t^{i,j}$ to $v_{t+1}^{i',j'}$, initiated by user $i \in N$ with private key index set $K^i$ and $j \in K^i$, receiver set $R = \{i' \in N\}$ with corresponding $j'\in K^{i'}$ at time $t$, is a state transition request represented by a \textit{tuple} $\mathcal{T}$ of the form
\begin{equation}\label{transition}
  \mathcal{T}_t^{i,i'} = \langle
  h_{n_t},
  pk_i^j,
  v_{t-1}^{\tilde{i},\tilde{j}},
  v_{t}^{i,j},
  v_{t+1}^{i',j'},
  s_i^j
  \rangle,
\end{equation}
where $s_i^j$ is a signature generated by $sk_i^j$, $v_{t-1}^{\tilde{i},\tilde{j}}$ is the previous state of the voting power unit in question, $v_{t+1}^{i',j'}$ was either calculated by $i$, if $i'$ provided them with both $h_{n_{t+1}}$ and $h_{p_{t+1}}$ for a delegation, or by $i'$ if it is an irreversible transfer.
\end{definition}

\begin{definition}
  A reversal at time $t > 0$ is a state transition request where a voting power unit $v_{\tilde{t} + 1}^{i',j'}$ which was the output of $\mathcal{T}_{\tilde{t}}^{i, i'}$ at state time $\tilde{t} < t$ is mutated by $i$, who issues the following tuple:
  \begin{equation}\label{reversal}
    \mathcal{R}_t^{i, i', \tilde{t}} = \langle
  h_{n_{\tilde{t} + 1}},
  h_{p_{\tilde{t} + 1}},
  pk_i^j,
  v_{\tilde{t}}^{i,j},
  v_{\tilde{t}+1}^{i',j'},
  v_{t+1}^{i^{''},j^{''}},
  s_i^j
  \rangle,
  \end{equation}
where $i^{''}$ is some other voter in $N$ or a newly generated $sk_{i^{''}}^{j^{''}}$ by $i$.
\end{definition}

If we assume that the function $H$ is common-knowledge, anyone can calculate $v_t^{i,j}$ from the first three elements of \ref{transition}, but they cannot know who the receiver $i'$ is, because it is infeasible to calculate the arguments that resulted in $\tilde{v}_{t+1}^{i',j'}$. These could be $h_{n_{t+1}}$ and $h_{p_{t+1}}$ or it could even be the case that $\mathcal{T}_t^{i,i'}$ is the transition from $v_0^{i,j}$ to $v_2^{i^{''},j^{''}}$ in Figure \ref{fig:unit} and $i^{''} \equiv i'$ privately generated all $n_1$, $n_2$, $sk_{i'}^{j'}$ and $sk_{i''}^{j''}$.
This means that, in case of a transfer, the sensitive information of the receiver remains secret even if the transition was transmitted through a compromised channel. In the case of a delegation, a man in the middle of the peer-to-peer transfer of $h_{n_{t+1}}$ and $h_{p_{t+1}}$ will know the first two elements of \ref{reversal}, but maybe not the real identity $pk_{i',j'}$ of the receiver, but, since they do not know $sk_i^j$ nor $sk_{i'}^{j'}$, they cannot generate a valid $s_i^j$ so they cannot do anything catastrophic with this information.

The state $S_t = \{v_t^{i, j}; j \in K^i\}_{i \in N}$ at time $t \geq 0$,
which consists of all the voting power unit identifiers in use,
can be recreated along with its evolution by anyone who will follow the execution of the valid state transitions stored in an append-only \textit{public ledger}
\begin{equation}\label{ledger}
  L_T = \left\{
  \mathcal{T}_t^{i, i'},
  \mathcal{R}_t^{i, i', \tilde{t}} \Big|
  \{i, i'\} \subseteq N,
  i \neq i',
  \tilde{t} < t
  \right\}_{t = 0}^T \text{ where $T \geq 0$}.
\end{equation}

If we assume that \ref{ledger} is common knowledge and state transitions follow \ref{transition} and \ref{reversal}, any third-party validator can verify that an input unit $v_t^{i,j}$ of a state transition, which was must have been the output of another state transition at time $t' < t$, can be reconstructed by the information the sender provides, namely $h_{n_t}$, $pk_i^j$ and $v_{t'}^{\tilde{i}, \tilde{j}}$ and check that the accompanying signature $s_i^j$ was indeed generated by $sk_i^j$.
If we refer to Figure \ref{fig:unit} as an example, at $t=0$, the voting power unit $v_0^{i,j}$ belongs to $i$ because only this individual knows the hashes $H(n_0)$ and $H(pk_i^j)$, since this is the only individual who also knows the plain text $n_0$ and $pk_i^j$. At $t = 1$, if we assume that the unit is now $v_1^{i',j'}$, the individual $i$ can only return to the previous state $v_0^{i,j}$ if they can reveal both $H(n_1)$ and $H(pk_{i'}^{j'})$, which they cannot, even if they knew $\tilde{v}_1^{i',j'}$.
In a voting scenario, only $i'$ now knows both $n_1$ and $pk_{i'}^{j'}$.
In all cases, the revealed $pk$ will be used to verify a signature of the transition's elements, made by its corresponding secret key.

The existence of an initial state filled with a static number of units which are recycled or the ``genesis'' of new units during the process is out of scope for the current paper.
When a unit does not have an assigned owner, its identifier is $v_{-1}^{\tilde{i}, \tilde{j}}$ and the output unit is the equivalent of $v_0^{i, j}$ for an initial owner $i$.
The generation of these units can be assigned to \textit{registration tellers} as in \cite{clarkson2008civitas} or it can be the result of a successful execution of a consensus protocol as in byzantine fault tolerant state machines \cite{nakamoto2008peer}.

\section{Voting}

Assume that the voter options are represented by public keys which are common knowledge, whose secret keys are controlled by the entities behind them.
Assume also that when state validators come across transitions issued by these public keys, they discard them from the next state.
Finally, assume that each entity has an automated mechanism which generates transfer data for their voters.

When a voter wants to vote on an option with some of their voting power, they ask for transfer data from it's entity and then proceed with the issuing of the voting power transfer. It's their choice if they want to engage in cardinal or ordinal rank voting, approval voting or plurality voting.
The last state contains the cast votes and maybe some unallocated voting power.
The entities group the votes in public. There can exist overlaps in votes if some entity is dishonest.
For the tally to end, since the public keys and the last state are common knowledge, each entity displays the plain-text nonces that correspond to their grouped votes. Every observer can now calculate the accumulated voting power of each entity.

\section{Drawbacks}

A drawback of our approach compared to established approaches of distributed systems with similar structure, such as cryptocurrencies that follow the UTXO model \cite{nakamoto2008peer}, is that in this system an individual needs to store more data than a mnemonic and their generated key pairs locally. They will now also need to store every pair of random nonce and public key that was used to generate their voting power unit identifiers if they ever want to use them again. This is not an insurmountable problem but rather an additional design step that needs to be implemented by whichever entity uses this model in practice in the future.

\section{Conclusion}

We established the use of a commitment scheme that will help us track the voting power of citizens in a society while preserving their privacy and offering coercion resistance. The design is flexible enough to allow for enhanced democratic techniques without sacrificing the initial goal, namely, to allow the citizens to keep their anonymity while external validators can safeguard the process from imposter-generated transitions. It is our hope that this system will become useful for the general population in its attempt to democratize more and more aspects of their collective decision making in the future.

\bibliography{private_votes}

\begin{thebibliography}{10}

\bibitem{bernhard2013towards}
David Bernhard, Stephan Neumann, and Melanie Volkamer.
\newblock Towards a practical cryptographic voting scheme based on malleable
  proofs.
\newblock In {\em E-Voting and Identify: 4th International Conference, Vote-ID
  2013, Guildford, UK, July 17-19, 2013. Proceedings 4}, pages 176--192.
  Springer, 2013.

\bibitem{blum2016liquid}
Christian Blum and Christina~Isabel Zuber.
\newblock Liquid democracy: Potentials, problems, and perspectives.
\newblock {\em Journal of Political Philosophy}, 24(2):162--182, 2016.

\bibitem{chase2012malleable}
Melissa Chase, Markulf Kohlweiss, Anna Lysyanskaya, and Sarah Meiklejohn.
\newblock Malleable proof systems and applications.
\newblock In {\em Advances in Cryptology--EUROCRYPT 2012: 31st Annual
  International Conference on the Theory and Applications of Cryptographic
  Techniques, Cambridge, UK, April 15-19, 2012. Proceedings 31}, pages
  281--300. Springer, 2012.

\bibitem{clarkson2008civitas}
Michael~R Clarkson, Stephen Chong, and Andrew~C Myers.
\newblock Civitas: Toward a secure voting system.
\newblock In {\em 2008 IEEE Symposium on Security and Privacy (sp 2008)}, pages
  354--368. IEEE, 2008.

\bibitem{ford2020liquid}
Bryan Ford.
\newblock A liquid perspective on democratic choice.
\newblock {\em arXiv preprint arXiv:2003.12393}, 2020.

\bibitem{grewal2013caveat}
Gurchetan~S Grewal, Mark~D Ryan, Sergiu Bursuc, and Peter~YA Ryan.
\newblock Caveat coercitor: Coercion-evidence in electronic voting.
\newblock In {\em 2013 IEEE Symposium on Security and Privacy}, pages 367--381.
  IEEE, 2013.

\bibitem{gutoski2015hierarchical}
Gus Gutoski and Douglas Stebila.
\newblock Hierarchical deterministic bitcoin wallets that tolerate key leakage.
\newblock In {\em Financial Cryptography and Data Security: 19th International
  Conference, FC 2015, San Juan, Puerto Rico, January 26-30, 2015, Revised
  Selected Papers 19}, pages 497--504. Springer, 2015.

\bibitem{juels2005coercion}
Ari Juels, Dario Catalano, and Markus Jakobsson.
\newblock Coercion-resistant electronic elections.
\newblock In {\em Proceedings of the 2005 ACM Workshop on Privacy in the
  Electronic Society}, pages 61--70, 2005.

\bibitem{karoukis2021deliberative}
Dimitrios Karoukis.
\newblock Deliberative democracy with dilutive voting power sharing.
\newblock {\em arXiv preprint arXiv:2109.01436}, 2021.

\bibitem{koblitz1987elliptic}
Neal Koblitz.
\newblock Elliptic curve cryptosystems.
\newblock {\em Mathematics of computation}, 48(177):203--209, 1987.

\bibitem{kulyk2017coercion}
Oksana Kulyk, Stephan Neumann, Karola Marky, Jurlind Budurushi, and Melanie
  Volkamer.
\newblock Coercion-resistant proxy voting.
\newblock {\em computers \& Security}, 71:88--99, 2017.

\bibitem{maxwell2011deterministic}
Gregory Maxwell and Iddo Bentov.
\newblock Deterministic wallets, 2011.

\bibitem{merkle1987digital}
Ralph~C Merkle.
\newblock A digital signature based on a conventional encryption function.
\newblock In {\em Conference on the theory and application of cryptographic
  techniques}, pages 369--378. Springer, 1987.

\bibitem{nakamoto2008peer}
Satoshi Nakamoto and A~Bitcoin.
\newblock A peer-to-peer electronic cash system.
\newblock {\em Bitcoin.--URL: https://bitcoin. org/bitcoin. pdf}, 4, 2008.

\bibitem{nejadgholi2021short}
Mahdi Nejadgholi, Nan Yang, and Jeremy Clark.
\newblock Short paper: ballot secrecy for liquid democracy.
\newblock In {\em Financial Cryptography and Data Security. FC 2021
  International Workshops: CoDecFin, DeFi, VOTING, and WTSC, Virtual Event,
  March 5, 2021, Revised Selected Papers 25}, pages 306--314. Springer, 2021.

\bibitem{smyth2019athena}
Ben Smyth.
\newblock Athena: A verifiable, coercion-resistant voting system with linear
  complexity.
\newblock {\em Cryptology ePrint Archive}, 2019.

\bibitem{spycher2012new}
Oliver Spycher, Reto Koenig, Rolf Haenni, and Michael Schl{\"a}pfer.
\newblock A new approach towards coercion-resistant remote e-voting in linear
  time.
\newblock In {\em Financial Cryptography and Data Security: 15th International
  Conference, FC 2011, Gros Islet, St. Lucia, February 28-March 4, 2011,
  Revised Selected Papers 15}, pages 182--189. Springer, 2012.

\bibitem{zhang2019statement}
Bingsheng Zhang and Hong-Sheng Zhou.
\newblock Statement voting.
\newblock In {\em Financial Cryptography and Data Security: 23rd International
  Conference, FC 2019, Frigate Bay, St. Kitts and Nevis, February 18--22, 2019,
  Revised Selected Papers 23}, pages 667--685. Springer, 2019.

\bibitem{zhang2021tracking}
Yuzhe Zhang and Davide Grossi.
\newblock Tracking truth by weighting proxies in liquid democracy.
\newblock {\em arXiv preprint arXiv:2103.09081}, 2021.

\end{thebibliography}
\bibliographystyle{plain}
\end{document}